\documentclass[%
 reprint,
 amsmath,amssymb,
 aps,
]{revtex4-2}

\usepackage{graphicx}
\usepackage{dcolumn}
\usepackage{bm}
\usepackage{amsmath,mathrsfs,amssymb}
\usepackage{subfigure}
\usepackage{float}
\usepackage{xcolor}
\usepackage{booktabs} 
\usepackage{adjustbox} 
\usepackage{array} 
\usepackage{siunitx}
\usepackage{hyperref}
\begin{document}

\preprint{APS/123-QED}

\title{Black hole echos reflect the phase transition and fluctuations in Hawking radiation}

\author{
  Tianqi Yue$^{1}$ and 
  Jin Wang$^{2,}$
}
\email{Corresponding author:jin.wang.1@stonybrook.edu}
\affiliation{
  $^{1}$College of Physics, Jilin University, Changchun 130022, China \\
  $^{2}$Department of Chemistry, and Department of Physics and Astronomy, 
  Stony Brook University, Stony Brook, New York 11794, U.S.A.
}


\begin{abstract}
Black hole phase transitions and Hawking radiation are fundamental yet elusive phenomena. By combining stochastic dynamics on the generalized free energy landscape with semiclassical radiation, we show that their interplay generates a distinctive dynamical echo—a resonant correlation peak in the switching statistics between metastable black hole phases. This echo provides a dual signature, simultaneously probing microscopic quantum fluctuations and macroscopic transition kinetics. The mechanism is general and is demonstrated to be realizable in acoustic black hole analogues, offering a path toward laboratory observation of emergent black hole thermodynamics.
\end{abstract}

\maketitle

The reconciliation of general relativity and quantum mechanics remains a central goal of physics, with the emergence of spacetime from quantum degrees of freedom being a pivotal concept\cite{jacobson1995thermodynamics,Witten:1998qj,J.M.Maldacena:1999}. Black hole is an emergent object from strong gravitational interactions. Black holes are found to be thermodynamic objects through Bekenstein entropy and Hawking radiations with finite temperature  \cite{hawking1975particle,bekenstein1973black}. Black holes can have thermodynamic phases and phase transitions, including the celebrated Hawking-Page transition in AdS spacetime \cite{hawking1983thermodynamics}. The RN-AdS black hole, whose phase structure mirrors a van der Waals fluid, provides a fundamental model where this emergence can be quantitatively studied\cite{Kastor2009,Dolan2011,Dolan2011Pressure,Kubiznak2012}. This implies the advantage of the stationary asymptotic observer—a choice that circumvents the complexities of local measurements—to define a universal thermodynamic clock.

From a cosmological perspective, the transient dynamics of black hole phase transitions occur over formidable timescales, necessitating the study of non-equilibrium processes.This has motivated kinetic approaches \cite{Li2020,li2020thermal,li2022generalized,yang2022kinetics} that are based on the generalized free energy landscape.This landscape, which provides the effective potential to model evolution as a stochastic process along an order parameter, derives from the Euclidean gravitational path integral formulation—the same first-principles approach that yields the standard black hole thermodynamics.Moreover, it extends the static saddle-point description to encompass dynamical pathways, a paradigm rooted in Ginzburg-Landau theory of phase transition dynamics\cite{hohenberg1977theory}.

However, directly observing these transient dynamics or the subtle signatures of Hawking radiation remains an extreme experimental challenge. To circumvent this, we investigate evaporation dynamics where Hawking radiation intrinsically couples to background thermal fluctuations. This coupling gives rise to a distinct, computable signal: a dynamical echo. The echo serves as a novel theoretical probe, simultaneously encoding information about macroscopic thermodynamics (phase transitions) and microscopic quantum dynamics (Hawking radiation), providing a unique window into real-time black hole evolution.

\begin{figure*}[!ht]
    \centering
    \subfigure[]{
    \includegraphics[width=0.2\linewidth]{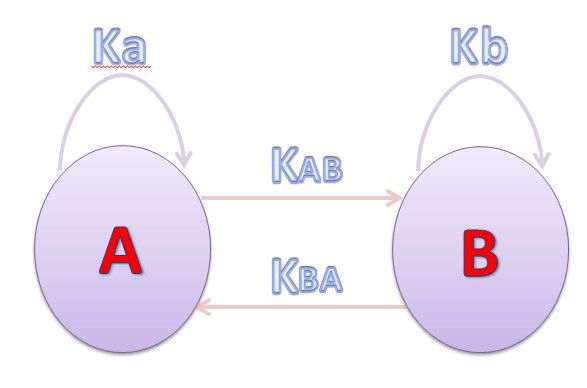}\label{fig:Full_Reaction}
    }
    \subfigure[]{
    \includegraphics[width=0.15\linewidth]{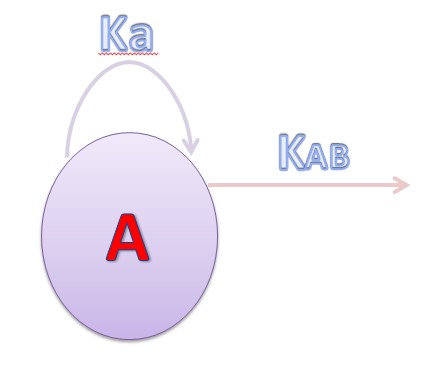} \label{fig:halfreactionA}
    }
    \subfigure[]{
    \includegraphics[width=0.15\linewidth]{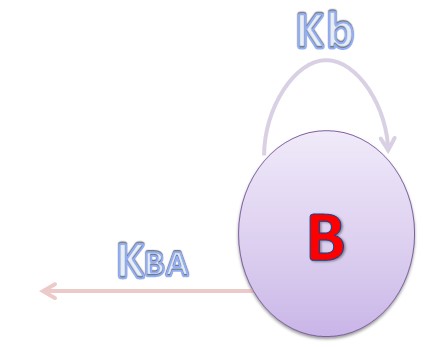 } \label{fig:halfreactionB}
    }
     \subfigure[]{
    \includegraphics[width=0.16\linewidth]{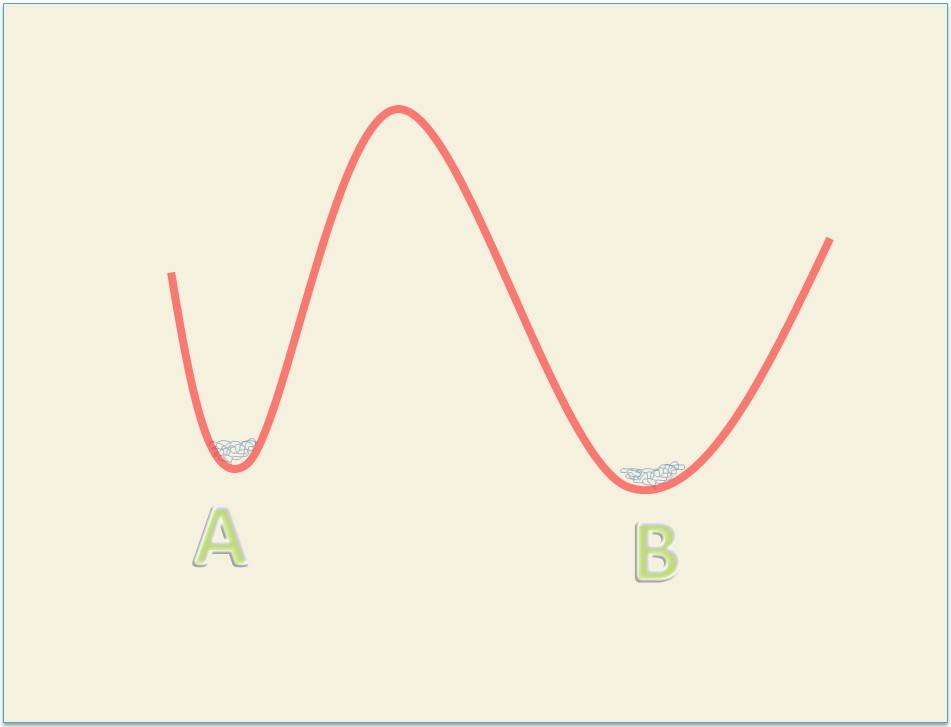}\label{fig:landscapewithreactionpath}
    }
    \subfigure[]{
    \includegraphics[width=0.16\linewidth]{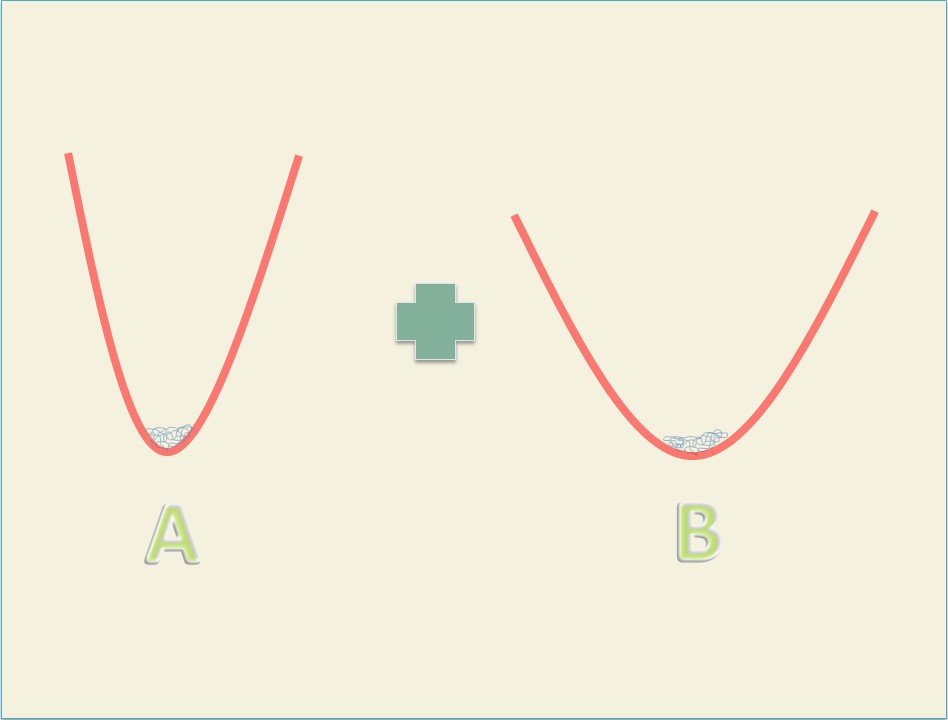} \label{fig:locallanscape}
    }
    \caption{ Conceptual framework. (a) Full phase transition network between small (A) and large (B) black hole phases. (b,c) Half-reactions: evaporation/absorption within each phase. (d,e) Free energy landscape \(G(r)\) showing the two basins (minima) separated by a barrier, with stochastic trajectories illustrating thermal fluctuations.}
\end{figure*}

The thermodynamic description of black holes originates from the Euclidean gravitational path integral formulism, where the partition function is dominated by classical, on-shell saddle points—the stationary black hole geometries. For the Reissner-Nordström anti-de Sitter (RN-AdS) spacetime, the on-shell metric is given by:
\begin{equation}
ds^2 = -f(r)dt^2 + f(r)^{-1}dr^2 + r^2 d\Omega^2
\end{equation}
where \(f(r) = 1 - \frac{2M}{r} + \frac{Q^2}{r^2} + \frac{r^2}{L^2}\).\(M\) is the mass, \( Q \) is the charge, and \( L =\sqrt{-3/\Lambda} \) is the AdS curvature radius (\( \Lambda\) being the cosmological constant). The thermodynamic pressure is related to the cosmological constant or AdS curvature radius \cite{Kastor2009, Dolan2011, Dolan2011Pressure}: \(P = \frac{3}{8 \pi} \frac{1}{L^2}\).

However,to describe the dynamics of phase transition, we need to describe not only the states, but also the process. This is realized by introducing a generalized free energy landscape \cite{li2022generalized}, which extends the concept of free energy to include intermediate transient states. A cornerstone of our framework is that this generalized landscape is fundamentally adapted to a stationary asymptotic observer. The time coordinate \(t\) in the metric serves as this observer's clock. This choice is physically imperative: quantum field theory, such as the Unruh effect \cite{unruh1976notes}, demonstrates that thermal states are observer-dependent. The asymptotic frame provides a unique, unambiguous reference where the Hawking temperature and black hole thermodynamics are unequivocally defined, avoiding local thermal ambiguities.

To capture the dynamics of fluctuating black holes away from equilibrium, one must consider off-shell configurations. This is achieved within the Euclidean gravitational path integral framework by evaluating the gravitational action on a broader class of geometries. The key object is the Einstein-Hilbert action with a cosmological constant:
\begin{equation}
I_E = -\frac{1}{16\pi} \int_{\mathcal{M}} \left( \mathcal{R} + \frac{6}{L^2} \right) \sqrt{g} \, d^4x,
\label{eq:EH_action}
\end{equation}
supplemented by boundary terms. For a black hole with an arbitrary Euclidean time period \(\beta = 1/T\) (not necessarily equal to the inverse Hawking temperature \(\beta_H\)), the Euclidean geometry develops a conical singularity at the horizon \(r = r_+\), with a deficit angle \(2\pi(1 - \beta/\beta_H)\). The action \(I_E\) evaluated on such a singular instanton receives an extra contribution proportional to the horizon area times this deficit angle\cite{li2022generalized}.

To obtain a finite result in asymptotically AdS space, one employs the background subtraction method: the divergent volume contribution of the black hole spacetime is regularized by subtracting the action of pure AdS space with a matched boundary geometry. Remarkably, after this subtraction and in the limit of a large cutoff radius, the total finite Euclidean action for a fluctuating RN-AdS black hole with charge \(Q\) and horizon radius \(r_+\) takes the form:
\begin{equation}
I_E = \beta \left[ \frac{r_+}{2}\left(1 + \frac{r_+^2}{L^2} + \frac{Q^2}{r_+^2}\right) \right] - \pi r_+^2.
\label{eq:finite_action}
\end{equation}
The first term in brackets is identified as the black hole mass \(M(r_+)\), and the second term \(S = \pi r_+^2\) is the Bekenstein-Hawking entropy.

From the path integral perspective, the generalized free energy \(G\) is directly related to this effective action via \(G = I_E / \beta\). Thus, taking the horizon radius \(r_+\) as the order parameter and denoting it simply as \(r\) for convenience, we arrive at the generalized free energy landscape \cite{Li2020, li2020thermal, li2022generalized}:
\begin{equation}
G(r) = \frac{I_E}{\beta} = \frac{r}{2}\left(1+\frac{r^2}{L^2}+\frac{Q^2}{r^2}\right) - \pi T r^2.
\label{G}
\end{equation}
This expression, emerging naturally from the Euclidean gravitational path integral, serves as the effective potential governing the stochastic dynamics of the black hole phase transition. It incorporates both the gravitational energy and the entropic contribution \( -TS \), with the temperature \(T\) now being an independent ensemble temperature that controls the landscape's shape.

The construction of \(G(r_+)\) can thus be viewed as a coarse-graining procedure from this privileged viewpoint, where microscopic degrees of freedom are integrated out to yield an effective potential for the order parameter. This landscape exhibits a double-basin structure, a hallmark of first-order phase transitions that directly mirrors the liquid-gas transition in a van der Waals fluid. This correspondence underscores universal, emergent thermodynamic principles.

To analyze the stochastic dynamics on this landscape, we employ the formalism of Ginzburg-Landau theory of phase transitions \cite{hohenberg1977theory}. This framework is directly applicable because the free energy landscape \(G(r)\) defines black hole metastable states and the barriers between them, which provides the essential input for a probabilistic description of stochastic transitions. We model the system as two distinct macrostates (two black hole phases)\cite{hanggi1990reaction,cao2000event,yang2001two}, each comprising numerous internal substates.This conceptual framework is illustrated in Figs.~\ref{fig:Full_Reaction}, \ref{fig:halfreactionA}, and \ref{fig:halfreactionB}.

The probability evolution over this network of states is governed by the master equation in its discrete, matrix form:
\begin{equation}
\begin{pmatrix}
   \dot{\rho}_a\\
   \dot{\rho}_b
\end{pmatrix}=
\begin{pmatrix}
   -K_a-K_{AB} & K_{BA}\\
   K_{AB} & -K_b-K_{BA}
\end{pmatrix}
\begin{pmatrix}
     \rho_a\\
   \rho_b
\end{pmatrix}.
\label{master equation}
\end{equation}
Here, \(K_a\) and \(K_b\) are matrices governing transitions within the substates of macrostate A and B (two black hole phases), respectively, while \(K_{AB}\) and \(K_{BA}\) govern the transitions between the two macrostates (two black hole phases).

The vectors \(F_a\) and \(F_b\) represent \cite{yang2001two} the steady-state probability fluxes over the internal substates of macrostate A and B, respectively. They correspond to the normalized probability fluxes into each macrostate at steady state (e.g., \(F_a \propto K_{BA} \rho_b\)). The Green's functions for the half-reactions, \(G_a(t) = e^{(-K_a - K_{AB})t}\) and \(G_b(t) = e^{(-K_b - K_{BA})t}\), provide the dynamical kernels for evolution starting from these distributions. From these, we define the single-event distribution for switching from A to B as \(f_a(t) = \sum K_{AB} G_a(t) F_a\), and for B to A as \(f_b(t) = \sum K_{BA} G_b(t) F_b\). The joint distribution for consecutive transitions A\(\to\)B\(\to\)A is \(f_{ab}(t_1, t_2) = \sum K_{BA} G_b(t_2) K_{AB} G_a(t_1) F_a\).

The dynamical echo is quantified by the difference function, which measures the correlation between a single switching event and two consecutive ones:
\begin{equation}
    \delta_{ab}(t_1,t_2) = \left|f_{ab}(t_1,t_2) - f_a(t_1) f_b(t_2)\right|.
    \label{delta}
\end{equation}
This function \(\delta\) captures the memory effect in the sequence \(A \xrightarrow{G_a(t_1)} B \xrightarrow{G_b(t_2)} A\). A pronounced maximum in \(\delta\)---the echo time---signals a strong causal correlation where the first transition predisposes the system to a subsequent reversal.

Our goal is to uncover how the stochastic phase transition dynamics and its echo imprint onto Hawking radiation.

We map the discrete matrix description to a continuous reaction-diffusion framework on the free energy landscape of a black hole (the small and large black hole basins, labeled A and B). The correspondence is: (1) The discrete intra-state transition matrix \(K_a\) becomes a continuous diffusion operator \(\hat{L}_{D_a}\) governing thermal fluctuations within a basin. (2) The transition rate \(K_{AB}\) from A to B remains as a constant term representing phase transitions. (3) We introduce the Hawking radiation rate \(K_{HR}(x)\) as a position-dependent reaction term specific to black holes, coupling evaporation to the local horizon radius.

This yields the continuous half-reaction equations specific to the black hole system. The landscape (Figs.~\ref{fig:landscapewithreactionpath} and \ref{fig:locallanscape}) determines both the diffusion (\(\hat{L}_{D_a}\)) and the phase transition rates via transition state theory \cite{Li2020,li2020thermal,li2022generalized}:
\[
K_{AB} = \frac{\omega_A \omega_{\text{max}}}{2\pi \eta} e^{-\beta \Delta G_{mA}},\qquad 
K_{BA} = \frac{\omega_B \omega_{\text{max}}}{2\pi \eta} e^{-\beta \Delta G_{mB}},
\]
where \(\omega_{A/B}\) are harmonic frequencies at the stable minima, \(\omega_{\text{max}}\) at the barrier top, \(\eta\) is the friction coefficient, \(\beta=1/(k_B T)\), and \(\Delta G_{mA/B}\) is the barrier height.

For a black hole in the small black hole phase (state A), the probability evolution is:
\begin{equation}
\begin{aligned}
     \frac{\partial \rho_a(x,t)}{\partial t} &= -K_{HR}(x) \rho_a(x,t) + \hat{L}_{D_a} \rho_a(x,t) \\
     &- K_{AB} \rho_a(x,t),
\end{aligned}
   \label{eq:main_half_reaction}
\end{equation}
where \(x = r - r_A\) is the deviation from the stable horizon radius \(r_A\), and the diffusion operator has the explicit form
\(
\hat{L}_{D_a} = \lambda\theta\frac{\partial}{\partial x}\left(\frac{\partial}{\partial x} + \frac{x}{\theta}\right),
\)
with \(\lambda\) being the thermal relaxation rate and \(\theta\) the equilibrium variance of fluctuations around the minimum. An analogous equation describes state B.

The conceptual foundation for Hawking evaporation in AdS spacetime was firmly established by Page, who demonstrated that under absorbing boundary conditions at infinity, an arbitrarily large black hole possesses a finite upper bound for its total decay time, scaling as \( t_d \sim l^{d-1}/G \) \cite{page2018finite}. This pivotal result confirms that AdS black holes are dynamic systems that can undergo complete evaporation.

Subsequent work \cite{li2021kinetics} pragmatically translated this physical picture into a tractable model by employing the Stefan-Boltzmann law to describe the mass-loss rate, capturing the essential coarse-grained dynamics.

Following this phenomenological approach, we use the Stefan-Boltzmann law (\(dM/dt \propto A T_H^4\)) to define the Hawking radiation rate for the stationary asymptotic observer, whose reference frame unambiguously defines the Hawking temperature and energy flux. From \cite{li2021kinetics}, this gives:
\begin{equation}
K_{HR} = \left|\frac{dM}{Mdt}\right|=\frac{ (1+8 \pi P r^2 - \frac{Q^2}{r^2})^4}{7680 \pi r^3 (1 + \frac{8 \pi Pr^2}{3 } + \frac{Q^2}{r^2})}.
\label{eq:main_HR_rate}
\end{equation}

A crucial consistency check confirms the validity of this approach: a straightforward dimensional analysis of Eq.~\eqref{eq:main_HR_rate} reveals that the characteristic evaporation time \( \tau_{\text{evap}} \sim 1/K_{HR} \) scales as \( \sim l^3 \) (in geometric units where \(G=1\)) for a large black hole (\( r_+ \sim l \)). This is in precise quantitative agreement with the finite lifetime \( t_d \sim l^3 \) derived by Page, demonstrating that the above expression correctly captures the core timescale physics of AdS black hole evaporation.

The Green's functions solving these equations allow us to construct the probability distributions for switching events. The echo is quantified by the difference function 
\[
\delta_{ba}(t_1, t_2) = K_{AB}K_{BA}\left| \Delta_{ba} - \Delta_b \Delta_a \right| e^{-K_{\text{eff}_b} t_1} e^{-K_{\text{eff}_a} t_2},
\] 
which measures the correlation between consecutive phase transitions. Here, the prefactor $\left| \Delta_{ba} - \Delta_b \Delta_a \right|$ captures the dynamical correlations between the two switching events, with $\Delta_{ba}$, $\Delta_a$, and $\Delta_b$ defined in terms of the Green's function parameters (see Appendix \ref{Green's Function} for their complete expressions \eqref{Pdeltaa},\eqref{Pdeltaba}). The effective decay rates $K_{\text{eff}_a} = K_{AB} + K_{a_1} + \lambda_a(s_a-1)/2$ and $K_{\text{eff}_b} = K_{BA} + K_{b_1} + \lambda_b(s_b-1)/2$ combine phase transition kinetics with Hawking radiation (through $K_{a_1}$, $K_{b_1}$) and thermal relaxation processes (through $\lambda_a$, $\lambda_b$, $s_a$, $s_b$). Our central finding is that $\delta_{ba}$ exhibits a clear maximum---the echo time---that is highly sensitive to the black hole parameters $(T, Q, P)$ and Hawking radiation, thus serving as a novel joint probe.

\begin{figure}[!ht]
    \centering
    \includegraphics[width=0.95\linewidth]{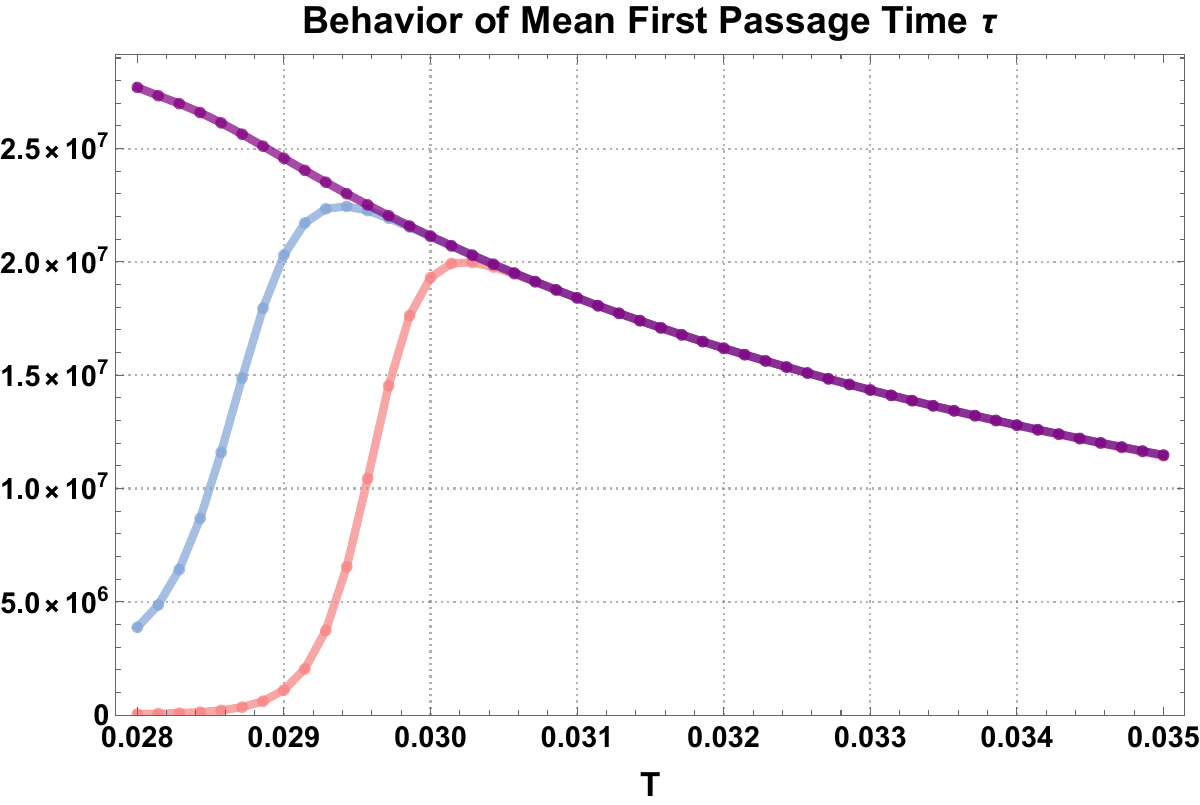}
    \caption{ Mean first-passage time (MFPT), calculated from the distribution \(f_{ba}(t,t)\), as a function of temperature. Curves correspond to dissipation coefficients \(\eta = 100\) (red), \(10^4\) (blue), and \(10^6\) (purple), at fixed \(Q=0.1\), \(P=0.003/(8\pi)\), and \(b_a=b_b=50\). }
    \label{fig:ScalingBehavior}
\end{figure}

\begin{figure*}[!ht]
    \centering
 
    \subfigure[]{
    \includegraphics[width=0.39\linewidth]{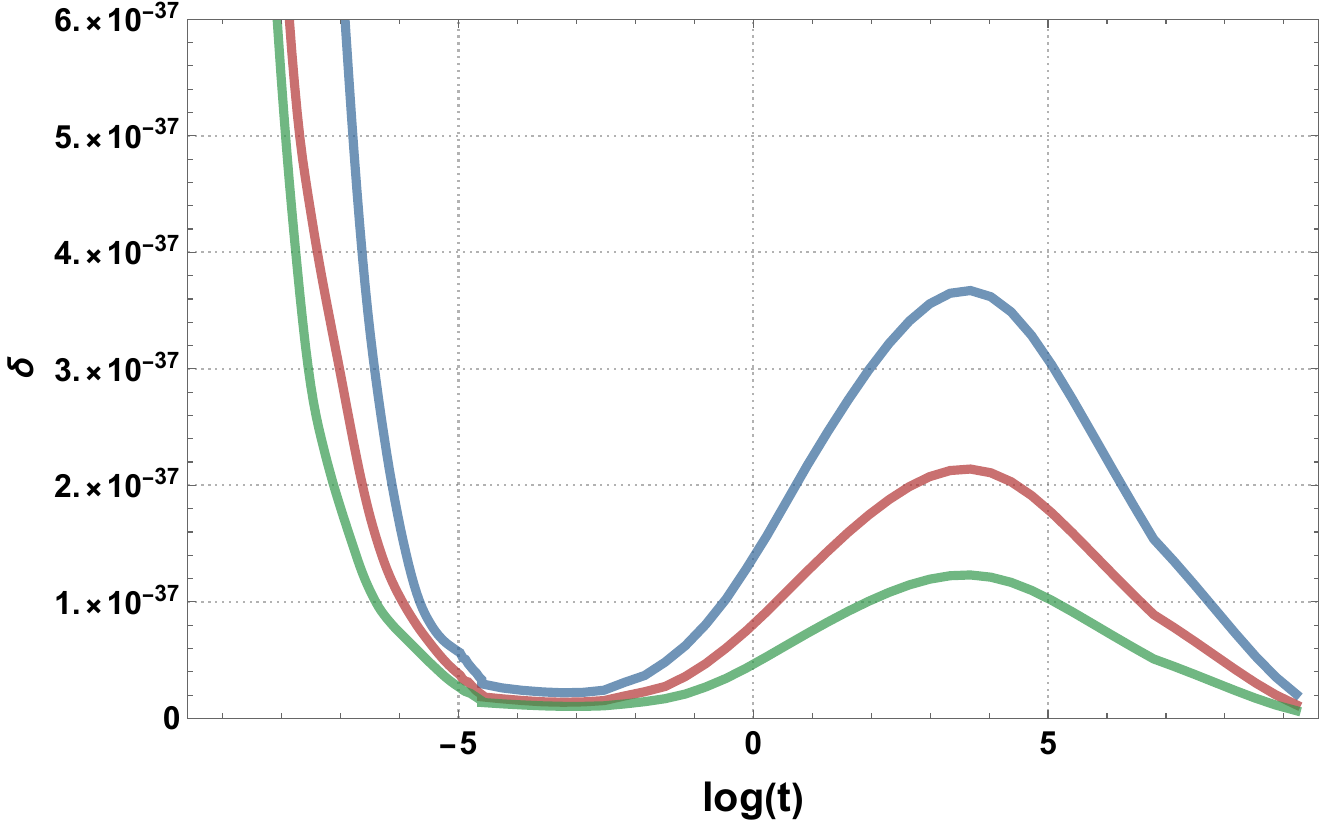}\label{fig:3RNecho}
    }
    \subfigure[]{
    \includegraphics[width=0.38\linewidth]{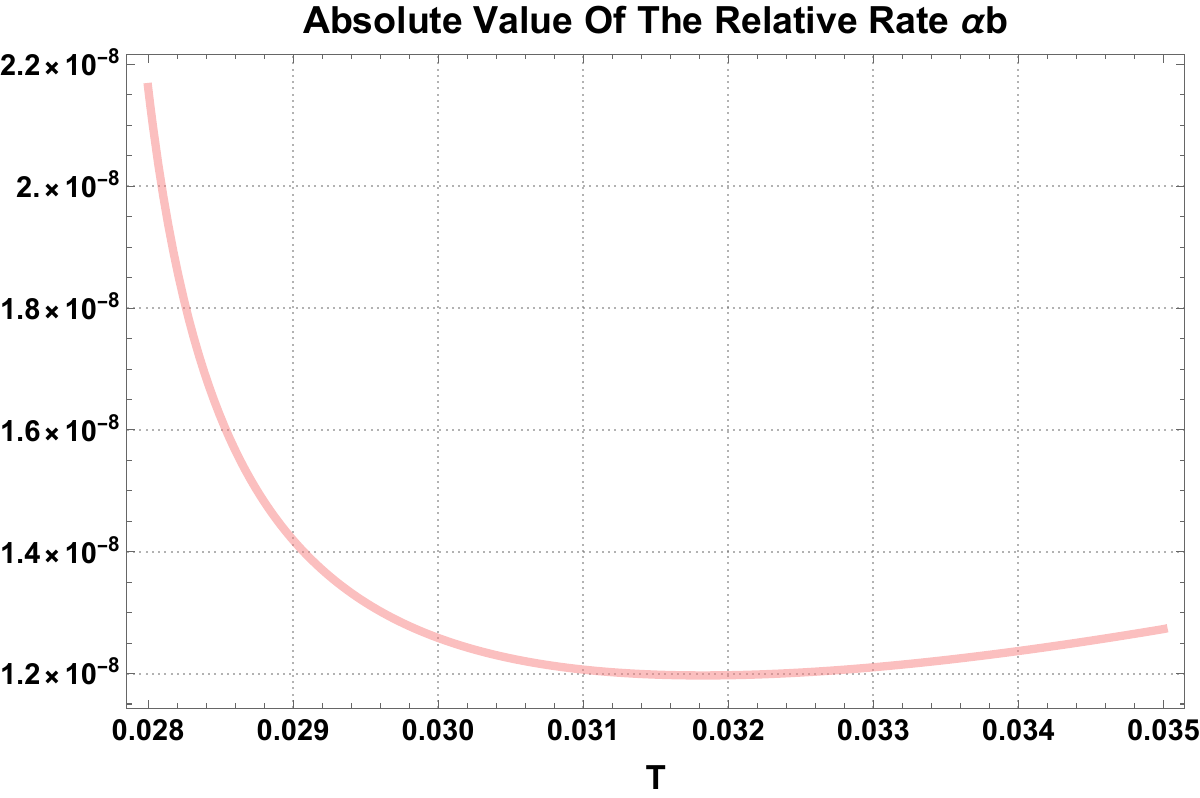}\label{fig:relativerate}
    }
    \caption{
 (a) Same-time difference function \(\delta(t)\) versus \(\log(t)\) for temperatures \(T = 0.0312\) (blue), \(0.0313\) (red), and \(0.0314\) (green), with fixed \(Q=0.1\), \(P=0.003/(8\pi)\), \(\eta=100\), and \(b_a=b_b=50\). The echo time is defined as the location of the maximum following an initial rapid decrease. (b) Relative Hawking radiation rate \( |\alpha_b|\) for the large black hole (state B) as a function of temperature.
}
    \label{fig:3}
\end{figure*}

\begin{figure*}[!ht]
    \centering
 
    \subfigure[]{
    \includegraphics[width=0.4585\linewidth]{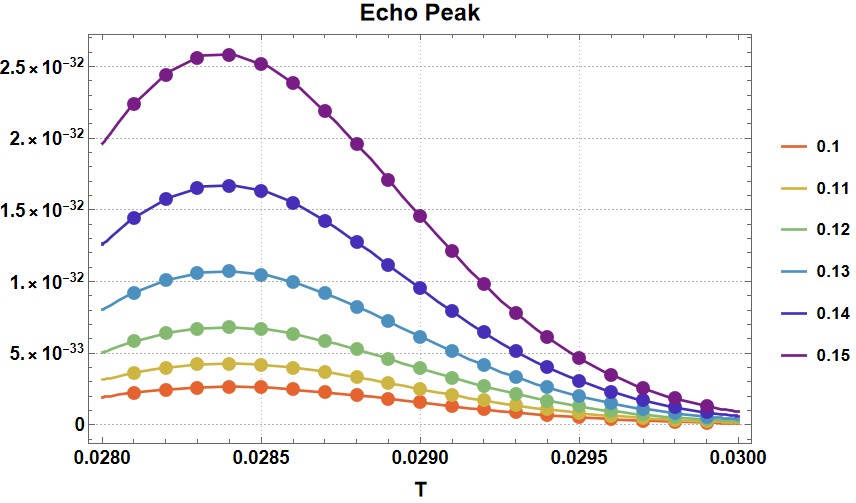}\label{fig:echopeakvsTandQ}
    }
    \subfigure[]{
    \includegraphics[width=0.5\linewidth]{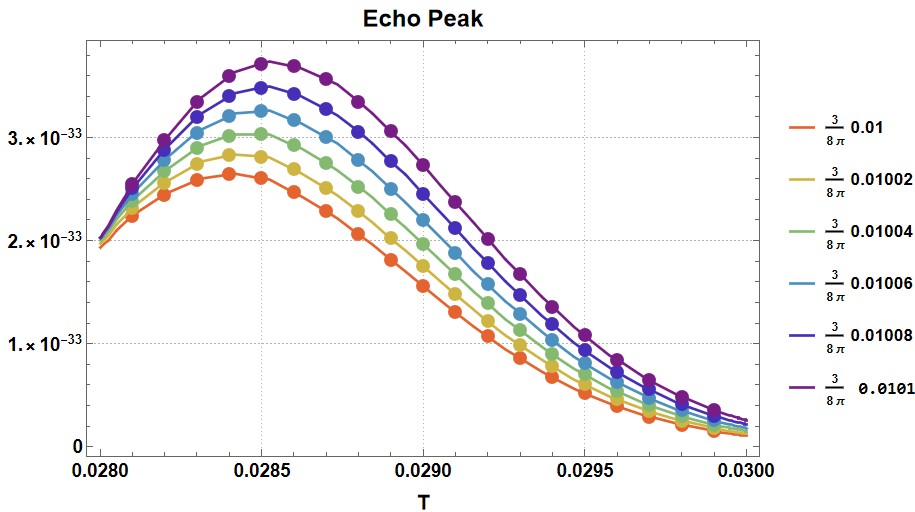}\label{fig:echopeakvsTandP}
    }
    \caption{Variation of the echo peak height with parameters. (a,b) The peak increases with \(Q \sim 0.1\text{--}0.15\) and \( P \sim \frac{3}{8 \pi}0.01\text{--}\frac{3}{8 \pi}0.0101 \), while showing a rise-and-fall behavior with \(T\), for fixed \(\eta=100\) and \(b_a=b_b=50\).
 }
    \label{fig:4}
\end{figure*}
\begin{table*}[!ht]
\caption{Echo peak amplitudes and echo times for different parameters with \(T=0.03,Q=0.1,P=\frac{3}{8 \pi} 0.01\).}
\label{tab:echo_data}
\begin{ruledtabular}
\begin{tabular}{@{}*{9}{c}@{}}
\toprule
$\eta$ & $b_a$ & $b_b$ & $K_{BA}$ & $K_{AB}$ & $K_{a_2}$ & $K_{b_2}$ & Echo Peak & Echo Time \\
\midrule
50    & 50    & 50  & $8.97394 \times 10^{-9}$  & $1.15006 \times 10^{-13}$ & $1.9637 \times 10^{-13}$ & $1.8728 \times 10^{-11}$ & $2.09332 \times 10^{-34}$ & 13.5656      \\
100   & 50    & 50  & $4.48697 \times 10^{-9}$  & $5.7503 \times 10^{-14}$ & $1.9637 \times 10^{-13}$ & $1.8728 \times 10^{-11}$ & $1.04431 \times 10^{-34}$ & 26.3525      \\
200   & 50    & 50  & $2.24349 \times 10^{-9}$  & $2.87515\times 10^{-14}$ & $1.9637 \times 10^{-13}$ & $1.8728 \times 10^{-11}$ & $5.22625 \times 10^{-35}$ & 52.0387     \\
100   & 100   & 50  & $4.48697 \times 10^{-9}$  & $5.7503 \times 10^{-14}$ & $2.45463 \times 10^{-14} $ & $1.8728 \times 10^{-11}$ & $1.35417 \times 10^{-35}$ & 30.3996      \\
100   & 50    & 100 & $4.48697 \times 10^{-9}$  & $5.7503 \times 10^{-14}$ & $1.9637 \times 10^{-13}$ & $2.341 \times 10^{-12}$ & $1.04384 \times 10^{-34}$ & 23.1872      \\
\bottomrule
\end{tabular}
\end{ruledtabular}
\end{table*}

The stochastic dynamics of black hole phase transitions gives rise to a distinct dynamical echo. This signal, quantified by the same-time difference function \(\delta(t) \equiv \delta_{ba}(t,t)\), provides a unique window into the interplay between the macroscopic thermodynamics of the free energy landscape and the microscopic dynamics tied to the spacetime geometry.

\textbf{Echo as a Probe of Hawking Radiation and Phase Transitions} The echo behavior emerges from a fundamental coupling between Hawking radiation and spacetime geometry. Gravitationally, Hawking radiation depends on curved spacetime; thermodynamically, this manifests as coupling between \(K_{HR}(x)\) and the horizon radius. This coupling strength is quantified by \(\alpha_a = (s_a - 1)/(4\theta_a)\) for phase A (and similarly \(\alpha_b = (s_b - 1)/(4\theta_b)\) for phase B), where \(s_a = \sqrt{1 - 4K_{a_2}\theta_a/\lambda_a}\) encodes the competition between Hawking radiation fluctuations (through \(K_{a_2}\)) and thermal relaxation (through \(\lambda_a\)), and \(\theta_a\) is the equilibrium variance of fluctuations around the minimum. Here \(K_{a_2} = K_{HR}(r_A)/(\sqrt{\pi}b_a^3)\) characterizes the spatial variation of \(K_{HR}(x)\) around the stable radius \(r_A\), with \(b_a\) being the width of the Gaussian smoothing function. Thus, \(\alpha_a\) measures the relative strength of Hawking radiation fluctuations compared to thermal relaxation—a competition that is central to the echo behavior (see Appendix \ref{Green's Function}).  

Crucially, when \(K_2 = 0\) (i.e., no spatial variation in \(K_{HR}(x)\) around the stable radius), we have \(s = \sqrt{1 - 4K_2\theta/\lambda} = 1\), which leads to \(\alpha = (s-1)/(4\theta) = 0\). In this case, the transformation \(G(x, y, t) = g(x, y, t) e^{\alpha(x^2 - y^2)}\) reduces to the identity transformation, and the Green's function simplifies to that of a standard Ornstein-Uhlenbeck process without the Hawking radiation coupling term. Consequently, the probability distributions become factorizable: the joint distribution factorizes into the product of single-event distributions, resulting in \(\Delta_{ba} = \Delta_b= \Delta_a=1\). This factorization implies \(\Delta_{ba} - \Delta_b\Delta_a = 0\), and thus \(\delta_{ba}(t_1,t_2) = 0\). Physically, this demonstrates that the echo signal requires spatial variations in Hawking radiation (\(K_2 \neq 0\)) to create correlations between consecutive switching events; without such fluctuations, the two events become statistically independent.

\textbf{Echo Resonance and Dynamical Competition}
The competition between Hawking radiation and thermal processes manifests in multiple ways. Figure~\ref{fig:relativerate} shows how \(\alpha\) varies with temperature, reflecting the changing balance between radiation fluctuations and relaxation. Similarly, Figure~\ref{fig:ScalingBehavior} demonstrates through the mean first-passage time how dissipation \(\eta\) modulates the competition between phase transition kinetics and evaporation. Near the critical point (Fig.~\ref{fig:echopeakvsTandQ}), the amplitude peaks as the two phases approach equal stability, making \(K_{AB} \approx K_{BA}\). Simultaneously, the echo time shifts with temperature, reflecting changes in both phase transition barriers and evaporation rates.
\begin{figure}[ht]
    \centering
    \includegraphics[width=0.8\linewidth]{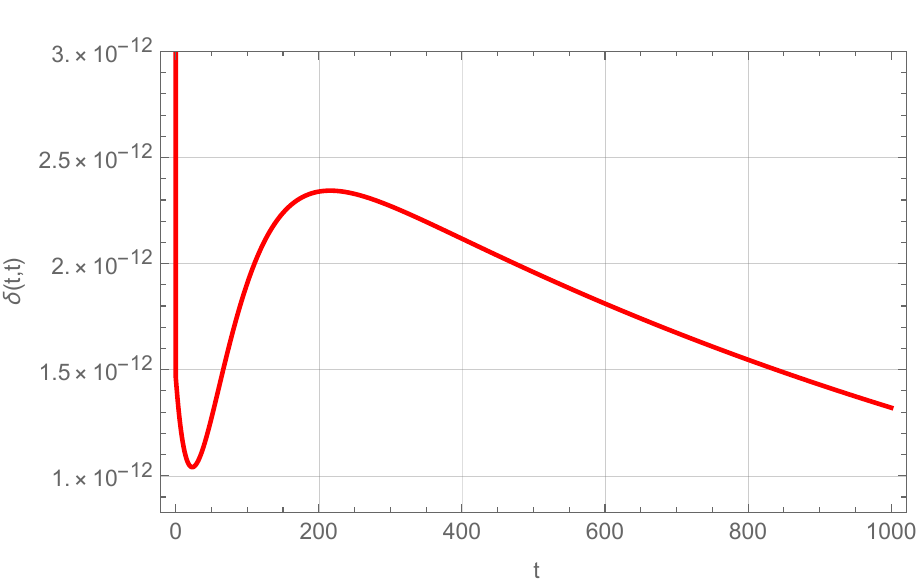}
    \caption{Numerical result for echo signals in analogue black hole phase transitions.  The correlation difference function \(\delta(t)\) reveals distinct echo peaks resulting from stochastic switching between the two phases. Parameters: \(A = 0.05\), \(L = 2\), \(J = 0.1\), \(T = 0.05\), \(m = 0.1\), \(g = 0.1\),\(K=1\)and \(K_{HR}=0.05 T_H^3\).}
    \label{fig:echo_signal}
\end{figure}

The echo amplitude provides information about Hawking radiation fluctuations through its dependence on \(\alpha\), while the echo time arises from the complex structure of \(\delta_{ba}(t_1,t_2)\), influenced by both exponential decay factors \(e^{-K_{eff}t}\) and prefactors that depend on \(\alpha\) and phase transition kinetics. This behavior extends to general time ratios \(k = t_1/t_2\) (Appendix~\ref{Dynamical Echo Function}).

\textbf{The distinct parameter dependence}
Table~\ref{tab:echo_data} reveals the distinct origins of these competitive effects: varying \(b\) (affecting Hawking radiation) primarily alters amplitude via \(\alpha\), while varying \(\eta\) (affecting phase transition rates) modifies both amplitude and echo time. Measuring both echo characteristics could thus provide constraints on the Hawking radiation-thermal relaxation competition (from amplitude variations) and phase transition kinetics (from echo time variations).

In conclusion, the echo behavior—rooted in the competition between Hawking radiation and thermal processes, with amplitude sensitive to radiation fluctuations and time encoding phase transition dynamics—provides a unique dual-probe of black hole non-equilibrium physics.

Echo signals originate from the interplay between phase transition kinetics and Hawking radiation fluctuations, providing a dynamical probe of black hole thermodynamics. Their amplitude reflects fluctuation strength while their timescale encodes transition rates. Numerical simulations based on the acoustic black hole model demonstrate clear echo signatures emerging from the stochastic switching between horizon configurations, as shown in Fig.~\ref{fig:echo_signal}.

These correlation dynamics may extend to other contexts such as gravitational waves from cosmological phase transitions\cite{hogan1986gravitational} and analogue gravity systems\cite{unruh1981experimental,weinfurtner2011measurement,kolobov2021observation,zhang2016thermodynamics,steinhauer2016observation,barcelo2001analogue}.

Using experimental parameters from the acoustic black hole analogy, as realized in Bose-Einstein condensate analogues \cite{steinhauer2016observation}, the effective Hawking evaporation rate is estimated as \(K_{HR} \sim \alpha^4 c_{\text{out}}/L\). Here, \(c_{\text{out}}\) is the speed of sound just outside the analogue event horizon, \(L\) represents the characteristic size of the analogue black hole, and \(\alpha\) is a dimensionless coupling parameter characterizing the interaction strength in the condensate. This rate reaches \(\mathcal{O}(0.1)\,\mathrm{s}^{-1}\) for typical configurations where the Hawking temperature satisfies \(k_{B}T_{H} \sim \alpha\, m c_{\text{out}}^2\) (with \(m\) representing the mass of an atom in the condensate), suggesting potential observability of echo-like correlations. 

The key to realizing such echoes lies in engineering the free energy landscape for horizon dynamics via a multi-extremum external potential \(V_{\text{ext}}(x)\). The sinusoidal profile \(V_{\text{ext}}(x) = A\sin(2\pi x/L + \phi)\) provides a concrete realization with alternating potential gradients. Through the hydrodynamic equations, \(V_{\text{ext}}(x)\) shapes the density \(\rho_0(x)\), which sets the horizon position \(x_h\) (where the flow velocity equals  local sound speed :\(|v_0| = c_s\)) and the free energy components \(E(x_h)\) and \(S(x_h)\) \cite{zhang2016thermodynamics} (detials in  Appendix\ref{A5}). 

The energy \(E(x_h)\) from Eq.~\eqref{eq:energy_derived} integrates contributions from \(dV_{\text{ext}}/dx\) and \(d\rho_0/dx\). The alternating gradient of \(V_{\text{ext}}(x)\) makes \(E(x_h)\) non-monotonic, while the entropy \(S(x_h) \propto \int^{x_h} \rho_0(x) dx\) grows monotonically. Their competition, tuned by temperature \(T\), creates double wells in \(F(x_h) = E(x_h) - T S(x_h)\), each minimum representing a distinct analogue black hole phase. Hawking radiation follows the Stefan-Boltzmann law for effective \((d+1)\)-dimensional geometry, giving radiation power \(P \propto T_H^3\) with \(T_H(x_h)\) from Eq.~\eqref{eq:hawking_temp}. Thus,evaporation is coupled directly to \(x_h\), linking phase transitions to quantum radiation. The stochastic switching between these phases produces a dynamical echo, numerically demonstrated in Fig.~\ref{fig:echo_signal}. This echo provides a direct signature of the Hawking radiation–phase transition coupling, offering a novel probe of black hole non-equilibrium thermodynamics.

For experimental implementation, creating the double-basin free energy landscape through tailored external potentials is essential. The resulting dynamical echo—emerging from the stochastic switching between the engineered phases—serves as a direct, observable signature of the intrinsic coupling between Hawking radiation and phase transition dynamics, offering a novel probe into black hole non-equilibrium thermodynamics.

\begin{acknowledgments}
T. Y. acknowledges support from the National Natural Science Foundation of China (Grant No. 12234019).
\end{acknowledgments}

\nocite{*}
\bibliography{apssamp}

\appendix
\section{Green's Function and Probability Distributions}\label{Green's Function}

\subsection{Derivation of the Green's Function}
The Green's function for the half-reaction equation satisfies:
\begin{equation}
\begin{aligned}
\frac{\partial}{\partial t} G(x, y, t) &= - (K_{a_1} + K_{AB})G(x, y, t) + K_{a_2}x^2 G(x, y, t) \\
&+ \lambda_a \theta_a \frac{\partial}{\partial x} \left( \frac{\partial}{\partial x} + \frac{x}{\theta_a} \right) G(x, y, t),
\end{aligned}
\end{equation}
with initial condition $G(x, y, 0) = \delta(x - y)$. Here $K_{a_1}$ and $K_{a_2}$ are expansion coefficients from Taylor expanding the product $K_{HR}(x)\delta_a(x)$, where $\delta_a(x) \approx e^{-x^2/b_a^2}/(\sqrt{\pi}b_a)$ is a Gaussian smoothing function with width $b_a$ that localizes the Hawking radiation effect around the horizon. Specifically, $K_{a_2} = K_{HR}(r_A)/(\sqrt{\pi}b_a^3)$.

Applying the transformation $G(x, y, t) = g(x, y, t) e^{\alpha_a(x^2 - y^2)}$, where $\alpha_a = (s_a - 1)/(4\theta_a)$ and $s_a = \sqrt{1 - 4K_{a_2}\theta_a/\lambda_a}$, reduces the equation to a standard Ornstein-Uhlenbeck form. The solution is:

\begin{equation}
\begin{aligned}
G_a(x,y,t) &= e^{-K_{\text{eff}_a} t} \left[ \frac{s_a}{2 \pi \theta_a (1 - e^{-2 \lambda_a s_a t})} \right]^{1/2} \\
&\times \exp\left[-\frac{s_a(x - y e^{-\lambda_a s_a t})^2}{2\theta_a (1 - e^{-2 \lambda_a s_a t})} + \alpha_a(x^2 - y^2)\right],
\end{aligned}
\end{equation}
where $K_{\text{eff}_a} = K_{AB} + K_{a_1} + \lambda_a(s_a-1)/2$.

\subsection{Probability Distributions for Switching Events}
The single-event distribution for switching from state A to B is:
\begin{equation}
f_a(t) = K_{AB} \iint G_a(x, y, t) \rho_b(y) \, dx\, dy = \Delta_a K_{AB} e^{-K_{\text{eff}_a} t},
\end{equation}
with $\rho_b(x) = e^{-x^2/(2\theta_b)}/\sqrt{2\pi\theta_b}$ and
\begin{equation}
\Delta_a = \sqrt{\frac{s_a}{4\theta_a\theta_b(1-e^{-2\lambda_a s_a t})}} \sqrt{\frac{1}{A_{a_1}B_{a_1}}},\label{Pdeltaa}
\end{equation}
where $A_{a_1} = B_a e^{-2\lambda_a s_a t} + 1/(2\theta_b) + \alpha_a$, $B_{a_1} = B_a - B_a^2 e^{-2\lambda_a s_a t}/A_{a_1} - \alpha_a$, and $B_a = s_a/[2\theta_a(1-e^{-2\lambda_a s_a t})]$.

The joint distribution for consecutive transitions A$\to$B$\to$A is:
\begin{equation}
\begin{aligned}
f_{ba}(t_1,t_2) = & K_{AB}K_{BA} \int_{-\infty}^{\infty} \int_{-\infty}^{\infty} \int_{-\infty}^{\infty} dx \, dy \, dz \\
& \times G_a(x, y, t_2) \, G_b(y, z, t_1) \, \rho_a(z) \\&=\Delta_{ba} K_{AB} K_{BA} e^{-K_{\text{eff}_b} t_1} e^{-K_{\text{eff}_a} t_2}.
\end{aligned}
\label{eq:appendix_fba}
\end{equation}

where
\begin{equation}
\begin{aligned}
\Delta_{ba} &= \sqrt{\frac{s_as_b}{8\theta_a^2\theta_b(1-e^{-2\lambda_a s_a t_2})(1-e^{-2\lambda_b s_b t_1})}} \sqrt{\frac{1}{A_{ba}B_{ba}C_{ba}}},\label{Pdeltaba} \\
A_{ba} &= B_b e^{-2\lambda_b s_b t_1} + \frac{1}{2\theta_a} + \alpha_b, \\
B_{ba} &= B_b - \frac{B_b^2 e^{-2\lambda_b s_b t_1}}{A_{ba}} + B_a e^{-2\lambda_a s_a t_2} + \alpha_a - \alpha_b, \\
C_{ba} &= B_a - \frac{B_a^2 e^{-2\lambda_a s_a t_2}}{B_{ba}} - \alpha_a,
\end{aligned}
\end{equation}
with $B_b = s_b/[2\theta_b(1-e^{-2\lambda_b s_b t_1})]$ and similar definitions for state B parameters.

\subsection{Dynamical Echo Function}\label{Dynamical Echo Function}
The echo is quantified by the difference function:
\begin{equation}
\delta_{ba}(t_1,t_2) = \left| f_{ba}(t_1,t_2) - f_b(t_1) f_a(t_2) \right|,
\end{equation}
which measures correlations between consecutive switching events. For equal times $t_1 = t_2 = t$, $\delta_{ba}(t,t)$ exhibits a characteristic maximum at the echo time, as shown in Fig. 3 of the main text. The full two-dimensional difference function $\delta_{ba}(t_1, t_2)$ provides complete characterization of echo behavior for different time ratios $t_1/t_2$.
\begin{figure}[!ht]
    \centering
    \subfigure[]{
    \includegraphics[width=0.99\linewidth]{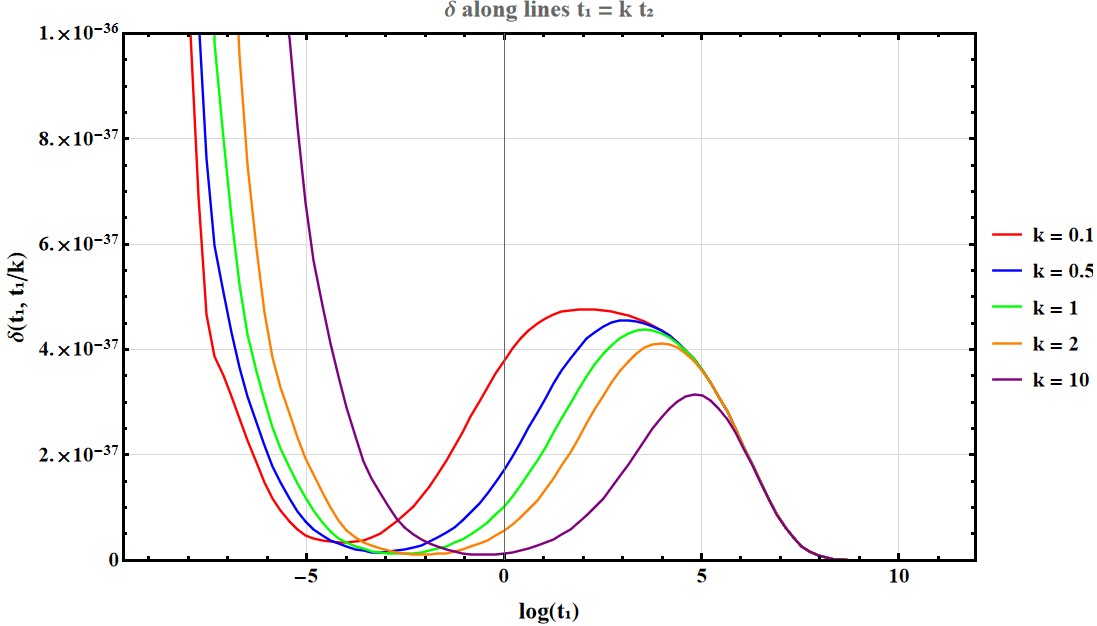} 
    } 
    \caption{ The difference function \(\delta(t_1, t_2/k)\) at fixed  parameters: \(T = 0.0312\), \(Q = 0.1\), \(P = 3/(8\pi) \times 0.01\),\(\eta=100\) and \(b_a=b_b=50\).}\label{fig:deltavsk}
\end{figure}

\section{Acoustic Black Hole Foundations}\label{A5}

\subsection{From Microscopic Action to Acoustic Metric}
Starting from the Gross-Pitaevskii action for a Bose-Einstein condensate (BEC)\cite{barcelo2001analogue},
\begin{equation}
S[\psi] = \int d^d x dt \left[ i\hbar\psi^*\partial_t\psi - \frac{\hbar^2}{2m}|\nabla\psi|^2 - V_{\text{ext}}|\psi|^2 - \frac{g}{2}|\psi|^4 \right],\label{eq:GPaction}
\end{equation}
we separate the field into background and fluctuations: $\psi = \psi_0 + \varphi$. Writing the background in Madelung form $\psi_0 = \sqrt{\rho_0}e^{i\theta_0}$ with density $\rho_0$ and phase $\theta_0$, and parameterizing fluctuations as $\varphi = \psi_0(\delta\rho/2\rho_0 + i\delta\theta)$, we obtain to quadratic order:
\begin{equation}
\begin{aligned}
    S_2[\delta\rho, \delta\theta] &= \frac{1}{2}\int d^d x \, dt \left[ -2\hbar \delta\rho \partial_t\delta\theta - \frac{\hbar^2}{4m\rho_0} (\nabla\delta\rho)^2 \right. \\
    &\left. - \frac{\hbar^2}{m} \rho_0 (\nabla\delta\theta)^2 - 2\hbar \delta\rho \mathbf{v}_0\cdot\nabla\delta\theta - g (\delta\rho)^2 \right],
\end{aligned}
\end{equation}
where $\mathbf{v}_0 = (\hbar/m)\nabla\theta_0$ is the background flow velocity. In the long-wavelength limit, defining $\phi = (\hbar/m)\delta\theta$ and eliminating $\delta\rho$ yields the effective phonon action:
\begin{equation}
S_{\text{eff}}[\phi] = \frac{1}{2}\int d^d x dt \, \rho_0 \left[ \frac{1}{c_s^2} (\partial_t\phi + \mathbf{v}_0\cdot\nabla\phi)^2 - (\nabla\phi)^2 \right],
\end{equation}
with $c_s = \sqrt{g\rho_0/m}$ the local speed of sound. This corresponds to propagation in the acoustic metric:
\begin{equation}
ds^2 = \frac{\rho_0}{m c_s} \left[ -c_s^2 dt^2 + (d\mathbf{x} - \mathbf{v}_0 dt)^2 \right].\label{B4}
\end{equation}

\subsection{Background Profile from Bernoulli Equation}
\label{A5:bernoulli}

The stationary background profile $\psi_0 = \sqrt{\rho_0}e^{i\theta_0}$ is obtained from the Euler-Lagrange equation of the action \eqref{eq:GPaction}. In the Thomas-Fermi approximation where quantum pressure is negligible, the background satisfies a Bernoulli-type equation:
\begin{equation}
g\rho_0^3 - [\mu - V_{\text{ext}}(x)]\rho_0^2 + \frac{mJ^2}{2} = 0,
\label{eq:background_cubic}
\end{equation}
where $\mu$ is the chemical potential and $J = \rho_0 v_0$ is the conserved mass current with $v_0 = (\hbar/m)\partial_x\theta_0$. Eq.~\eqref{eq:background_cubic} determines the density profile $\rho_0(x)$ for a given external potential $V_{\text{ext}}(x)$. Sonic horizons emerge at positions $x_h$ where the flow velocity equals the local sound speed, $|v_0(x_h)| = c_s(x_h)$.

\subsection{Hawking Radiation and Horizon Thermodynamics}
\label{A5:hawking}

For a stationary one-dimensional flow along $x$, we perform the coordinate transformation $d\tau = dt + [v_0/(c_s^2 - v_0^2)]dx$ on the metric \eqref{B4}, obtaining:
\begin{equation}
ds^2 = \frac{\rho_0}{m c_s} \left[ -(c_s^2 - v_0^2)d\tau^2 + \frac{c_s^2}{c_s^2 - v_0^2}dx^2 \right].\label{eq:pg_metric}
\end{equation}

At the analogue horizon $x_h$ where $|v_0(x_h)| = c_s(x_h)$, we expand to first order: $v_0(x) \approx -c_s(x_h) + v_0'(x_h) (x - x_h)$ and $c_s(x) \approx c_s(x_h) + c_s'(x_h)(x - x_h)$. Substituting into Eq.~\eqref{eq:pg_metric} yields the near-horizon Rindler form. Introducing the proper distance $\rho = \sqrt{2(x - x_h)/\kappa}$, where $\kappa = |v_0'(x_h) + c_s'(x_h)|$ is the surface gravity, the metric simplifies to $ds^2 \approx -\kappa^2 \rho^2 d\tau^2 + d\rho^2$, which describes a Euclidean Rindler space with temperature:
\begin{equation}
T_H = \frac{\hbar \kappa}{2\pi k_B} = \frac{\hbar}{2\pi k_B} \left| \frac{dv_0}{dx}(x_h) + \frac{dc_s}{dx}(x_h) \right|.
\label{eq:hawking_temp}
\end{equation}

\subsection{Free Energy Landscape for Phase Transitions}
\label{A5:free_energy}

The free energy landscape governing analogue black hole phase transitions can be derived from the steady-flow hydrodynamic equation\cite{zhang2016thermodynamics}:
\begin{equation}
\rho_0 v_0 \frac{dv_0}{dx} + \frac{\rho_0}{m} \frac{dV_{\text{ext}}}{dx} + c_s^2 \frac{d\rho_0}{dx} = 0.
\label{eq:euler_analogue}
\end{equation}

At the horizon $x_h$, using Eq.~\eqref{eq:hawking_temp} for the temperature and Eq.~\eqref{eq:euler_analogue}, we obtain the differential relation:
\begin{equation}
\begin{aligned}
  &\frac{\hbar}{2\pi} \left( v_0'(x_h) + c_s'(x_h) \right) \frac{2\pi}{\hbar K} \rho_0(x_h) dx \\&
- \left( \frac{\rho_0(x_h)}{m c_s(x_h) K} dV_{\text{ext}}(x_h) + \frac{3c_s(x_h)}{2K} d\rho_0(x_h) \right) = 0,  
\end{aligned}
\end{equation}
where $K$ is a constant with dimensions of length. This suggests the identification of entropy and energy differentials. Integrating from the system boundary to $x_h$ gives the horizon-dependent entropy and energy:
\begin{align}
S(x_h) &= \frac{2\pi}{\hbar K} \int_0^{x_h} \rho_0(x) dx, \label{eq:entropy_derived} \\
E(x_h) &= \frac{1}{K} \int_0^{x_h} \left[ \frac{\rho_0(x)}{m c_s(x)} \frac{dV_{\text{ext}}}{dx} + \frac{3c_s(x)}{2} \frac{d\rho_0}{dx} \right] dx. \label{eq:energy_derived}
\end{align}
The horizon position $x_h$ serves as the order parameter, and the free energy landscape is:
\begin{equation}
F(x_h, T) = E(x_h) - T S(x_h), \label{eq:free_energy_def}
\end{equation}
where $T$ is the ambient temperature. For potentials $V_{\text{ext}}(x) = A\sin(2\pi x/L + \varphi)$, $E(x_h)$ oscillates while $S(x_h)$ grows monotonically, enabling double-well structures and first-order phase transitions between ``small'' and ``large'' analogue black holes. The relative depth of the wells is tunable through $A$, $L$, $\varphi$, and $J$.

Free energy minima correspond to stable horizon configurations, while maxima represent transition states. Thermal fluctuations at finite $T$ enable stochastic switching between minima, with rates governed by Kramers theory.
\end{document}